\documentclass[conference]{IEEEtran}

\usepackage[dvips]{color}
\usepackage{graphicx}
\usepackage{amsmath}

\begin{document}

\title{Outage Analysis of Multi-Relay Selection for Cognitive Radio with Imperfect Spectrum Sensing} \normalsize

\author{
\IEEEauthorblockN{Yulong~Zou, Jia~Zhu, and Baoyu~Zheng}

\IEEEauthorblockA{School of Telecomm. \& Inform. Eng., Nanjing Univ. of Posts and Telecomm., Nanjing, China}
\IEEEauthorblockA{Email: \{yulong.zou, jiazhu, zby\}@njupt.edu.cn}

}

\maketitle

\begin{abstract}
In this paper, we examine the outage performance of a cognitive relay network, which is comprised of a secondary transmitter (ST), multiple decode-and-forward (DF) relays and a secondary destination (SD). We propose a multi-relay selection scheme for the cognitive relay network, where multiple relays are selected and used to participate in forwarding the secondary transmission from ST to SD. A closed-form expression of the outage probability for the proposed multi-relay selection under imperfect spectrum sensing is derived in Rayleigh fading environments. For comparison purposes, the conventional direct transmission and the best-relay selection are also considered as benchmarks. Numerical results show that as the spectrum sensing performance improves with an increasing detection probability and/or a decreasing false alarm probability, the outage probabilities of the proposed multi-relay selection as well as the direct transmission and the best-relay selection schemes all decrease accordingly. It is also demonstrated that the proposed multi-relay selection significantly outperforms the conventional approaches in terms of the outage probability.

\end{abstract}

\begin{IEEEkeywords}
Outage probability, relay selection, cognitive relay, cognitive radio, spectrum sensing.

\end{IEEEkeywords}

\IEEEpeerreviewmaketitle

\section{Introduction}

\IEEEPARstart Cognitive radio (CR) as a dynamic spectrum access (DSA) enabling technique allows unlicensed users (known as secondary users) to communicate with each other over the unused licensed bands (called spectrum holes) detected through spectrum sensing [1]. To be specific, secondary users first scan the licensed bands of interest through spectrum sensing for identifying spectrum holes [2] and then start to transmit over the detected spectrum holes (if any) [3]. Due to the background noise and wireless fading effects, the perfect spectrum sensing without any miss detection and false alarm is impossible. Moreover, the licensed bands may be occupied by licensed users (also referred to as primary users) and become unavailable for the secondary users. Therefore, the secondary transmission throughput is typically limited due to the imperfect spectrum sensing and the spectrum unavailability [4], [5].

Cooperative relaying is emerging as an effective means to improve the spectrum sensing and the secondary transmission of cognitive radio networks. In [6] and [7], it was shown that user cooperation can significantly enhance the spectrum sensing performance in fading environments in terms of the detection probability and false alarm probability, where a dedicated reporting channel is assumed when cooperative users transmit their initial sensing results to a fusion center. Later on, a selective-relay based cooperative sensing scheme without the dedicated reporting channel was proposed in [8], showing a significant sensing performance improvement with the aid of cooperative relays. In [9] and [10], we investigated the employment of cooperative relays for secondary transmissions in cognitive radio networks and showed the outage improvement of secondary transmissions using the best-relay selection. In [11], Kim \emph{et al.} proposed a full-channel state information (CSI)-based and a partial CSI-based relay selection schemes for a single-carrier spectrum sharing system. Additionally, in [12], Li and Nosratinia examined a distributed relay selection and clustering framework for a spectrum-sharing network.


Although there are extensive research efforts devoted to the relay selection for cognitive radio networks, most of existing works [9]-[12] are focused on the single best-relay selection. In this paper, we are motivated to explore multi-relay selection for a cognitive radio network, where a secondary transmitter (ST) is intended to transmit to a secondary destination (SD) with the help of multiple decode-and-forward (DF) relays. To be specific, ST first broadcasts its signal to the DF relays, which attempt to decode their received signals. Then, the DF relays of successfully decoding (which may be more than one) are chosen to forward ST's signal to SD. We derive a closed-form expression of the outage probability for the proposed multi-relay selection assisted secondary transmissions with imperfect spectrum sensing.



\section{Multi-Relay Selection in Cognitive Radio Networks}

\subsection{System Model}
Fig. 1 shows a primary network sharing its licensed spectrum with a secondary cognitive relay network, where the secondary network first scans the licensed spectrum for identifying spectrum holes and then access the detected holes for data transmissions. If no spectrum hole is found, ST is not allowed to transmit and keeps silent to avoid interfering with primary users (PUs). If a spectrum hole is identified, ST will start to transmit to SD with the aid of $N$ relays. Throughout this paper, the relays are assumed to operate with the decode-and-forward (DF) protocol and, moreover, similar performance results could be obtained for the amplify-and-forward (AF) protocol. For notational convenience, all the $N$ relays are denoted by ${\cal R}=\{{\textrm{R}}_1,{\textrm{R}}_2,\cdots,{\textrm{R}}_N\}$. Let $ {H_0}$ and $H_1$ represent the licensed spectrum being unoccupied and occupied by the primary transmitter (PT), respectively. Additionally, ${\hat H} $ is used to indicate the status of the licensed spectrum estimated by the secondary network.
\begin{figure}
  \centering
  {\includegraphics[scale=0.55]{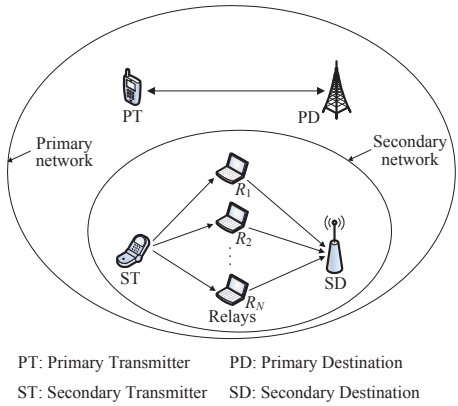}\\
  \caption{A primary network coexists with a secondary cognitive relay network.}\label{Fig1}}
\end{figure}

Due to the wireless fading and noise effects, there always exist the miss detection and false alarm of the presence of spectrum holes. Here, let $P_d$ and $P_f$ denote the probability of detection and false alarm of the presence of a spectrum hole, respectively, i.e., $P_d  = \Pr ( {\hat H  = H_0 |H_0 } ) $ and  $P_f  = \Pr ( {\hat H  = H_0 |H_1 } ) $. For the sake of guaranteeing that the interferences generated from ST and relays are tolerable at PUs, the detection probability ${{{P}}_{{d}}}$ and false alarm probability ${{{P}}_{{f}}}$ should be within a meaningful range. Besides, all the channels between any two network nodes of Fig. 1 are modeled as the Rayleigh fading. Finally, all receivers are assumed to have the zero-mean additive white Gaussian noise (AWGN) with a variance of $N_0$.

\subsection{Proposed Multi-Relay Selection}
This subsection proposes a multi-relay selection scheme for the cognitive relay network of Fig. 1, where ST can't directly communicate with SD and $N$ DF relays are available to assist the ST's transmission to SD. More specifically, if a spectrum hole is found, ST first broadcasts its signal to $N$ DF relays, part of which are then selected to forward their decode outcomes to SD. For notational convenience, the signals to be transmitted by ST and PT are denoted by $x_s$ and $x_p$, respectively, where $E[|x_p|^2]=E[|x_s|^2]=1$. Moreover, let $P_s$ and $P_p$ denote the transmit powers of the ST and PT, respectively. Given that a spectrum hole is identified by the secondary network (i.e. ${\hat H} = {H_0}$), ST starts to transmit its signal ${x_s}$ at a data rate of $R$. Thus, the received signal at a relay ${\textrm{R}}_i$ is written as
\begin{equation}\label{equa1}
y_i  = h_{si} \sqrt {P_s } x_s  + h_{pi} \sqrt {\alpha P_p } x_p  + n_i,
\end{equation}
where $h_{si}$ and $h_{pi}$ are the fading coefficients of the ST-R$_i$ channel and that of the PT-R$_i$ channel, respectively, $n_i$ is the AWGN experienced at R$_i$, and $\alpha$ is given by
\begin{equation}\label{equa2}
\alpha = \left\{ \begin{array}{l}
 0,\quad H_0 \\
 1,\quad H_1. \\
 \end{array} \right.
\end{equation}
By using (1), the capacity of the ST-R$_i$ channel can be obtained as
\begin{equation}\label{equa3}
C_{si}  = \frac{1}{2}\log _2 ( {1 + \frac{{|h_{si} |^2 \gamma _s }}{{\alpha |h_{pi} |^2 \gamma _p  + 1}}} ),
\end{equation}
where $\gamma_s=P_s/N_0$, $\gamma_p=P_p/N_0$, and the factor $\frac{1}{2}$ in the front of $\log(\cdot)$ is due to the fact that two time slots are consumed for transmitting the ST's signal $x_s$ to SD via R$_i$. All $N$ relays attempt to decode $x_s$ from their received signals. For notational convenience, let ${\cal {D}}$ denote the set of relays that successfully decode $x_s$. Given $N$ relays, there are $2^N$ possible combinations of the set ${\cal {D}}$, and thus the sample space of ${\cal {D}}$ can be expressed as $\Omega  = \left\{ {\emptyset ,{\cal {D}}_1 ,{\cal {D}}_2 , \cdots ,{\cal {D}}_n , \cdots ,{\cal {D}}_{2^N  - 1} } \right\}$, where $\emptyset$ is an empty set and $ {\cal {D}}_n$ is the $n{\textrm{-th}}$ non-empty subset of the $N$ relays. The well-known Shannon's coding theorem shows that when the channel capacity falls below the data rate, the receiver is deemed to fail to decode the source signal. Moreover, if the channel capacity becomes larger than the data rate, the receiver is able to succeed in decoding. Thus, using (3), the event of ${\cal {D}}=\emptyset$ can be described as
\begin{equation}\label{equa4}
C_{si}  < R,\quad i = 1,2, \cdots ,N.
\end{equation}
Meanwhile, the event of ${\cal {D}}={\cal {D}}_n$ is expressed as
\begin{equation}\label{equa5}
\begin{split}
&C_{si}  > R,\quad i \in {\cal {D}}_n  \\
&C_{sj}  < R,\quad j \in \bar {\cal {D}}_n,  \\
 \end{split}
\end{equation}
where $\bar {\cal {D}}_n={\cal {R}}-{\cal {D}}_n$ is the complementary set of ${\cal {D}}_n$. If $\cal {D}$ is an empty set, then all the relays transmit nothing and thus SD fails to decode $x_s$ in this case. If ${\cal {D}}$ is a non-empty set (e.g., ${\cal {D}}={\cal {D}}_n$), then the relays within ${\cal {D}}_n$ are employed for transmitting $x_s$ to SD. Note that the relays within the decoding set ${\cal {D}}_n$ are able to successfully decode $x_s$ and the others within $\bar {\cal {D}}_n$ fail to decode. To make an effective use of the multiple relays within ${\cal {D}}_n$, a weight vector as denoted by ${{\textbf{w}}} = [w_1 ,w_2 , \cdots ,w_{|{\cal{D}}_n|} ]^T$ is utilized at the relays for transmitting $x_s$, where $|{\cal {D}}_n|$ is the cardinality of ${\cal{D}}_n$. The weight vector ${{\textbf{w}}}$ has to be normalized according to $||{\textbf{w}}|| = 1$ for the sake of making the total transmit power at the relays being constrained to $P_s$. Hence, given ${\cal {D}}={\cal {D}}_n$ and considering that all the relays within ${\cal {D}}_n$ are selected for transmitting $x_s$ with the vector ${{\textbf{w}}}$, the received signal at SD can be expressed as
\begin{equation}\label{equa6}
y_d  = \sqrt {{P_s }} {\textbf{w}}^T {\textbf{H}}_d x_s  + \sqrt {\alpha P_p } h_{pd} x_p  + n_d,
\end{equation}
where ${\textbf{H}}_d  = [h_{1d} ,h_{2d} , \cdots ,h_{|{\cal{D}}_n |d} ]^T$. Assuming that ${\textbf{H}}_d$ is known at SD and using the coherent detection, we can obtain from (6) the signal-to-interference-and-noise ratios (SINR) at SD as
\begin{equation}\label{equa7}
{\textrm{SINR}}_d  = \frac{{\gamma _s }}{{\alpha |h_{pd} |^2 \gamma _p + 1}}|{\textbf{w}}^T {\textbf{H}}_d |^2.
\end{equation}
We aim to maximize the SINR ${\textrm{SINR}}_d$ through an optimization of the weight vector ${{\textbf{w}}}$ for improving the SD's capability of successfully decoding $x_s$, yielding
\begin{equation}\label{equa8}
\begin{split}
\mathop {\max }\limits_{\textbf{w}} {\textrm{SINR}}^{{\textrm{multi}}}_d ,\quad {\textrm{s.t. }}||{\textbf{w}}|| = 1.
\end{split}
\end{equation}
By using the Cauchy-Schwarz inequality, the optimal weight vector ${\textbf{w}}_{{\textrm{opt}}}$ can be readily obtained from (8) as
\begin{equation}\nonumber
{\textbf{w}}_{{\textrm{opt}}}  = \frac{{{\textbf{H}}_d^* }}{{|{\textbf{H}}_d |}},
\end{equation}
Substituting the optimal weight vector ${\textbf{w}}_{{\textrm{opt}}}$ into (7) yields
\begin{equation}\label{equa9}
{\textrm{SINR}}_d = \frac{{\gamma _s }}{{\alpha \gamma _p |h_{pd} |^2  + 1}}\sum\limits_{i \in {\cal{D}}_n } {|h_{id} |^2 }.
\end{equation}
By using the Shannon's channel capacity formula, the transmission capacity achieved at SD can be given by
\begin{equation}\label{equa10}
C_d = \frac{1}{2}\log _2 ( {1 + \frac{{\gamma _s }}{{\alpha \gamma _p |h_{pd} |^2  + 1}}\sum\limits_{i \in {\cal{D}}_n } {|h_{id} |^2 } } ),
\end{equation}
for the case of ${\cal{D}} = {\cal{D}}_n $.

\section{Outage Analysis of Proposed Multi-Relay Selection Scheme}
In this section, we present the outage probability analysis of proposed multi-relay selection scheme. As is known, an outage event occurs when the channel capacity drops below the data rate. Thus, given that a spectrum hole is identified, the outage probability of proposed multi-relay selection scheme is obtained from (10) as
\begin{equation}\label{equa11}
\begin{split}
P_{{\textrm{out}}}  = & \Pr ( {\left. {{\cal{D}} = \emptyset } \right|\hat H = H_0 } ) \\
&+ \sum\limits_{n = 1}^{2^N  - 1} {\Pr ( { {C_d  < R,{\cal{D}} = {\cal{D}}_n } |\hat H = H_0 } )},
\end{split}
\end{equation}
where $C_{{d}}$ is given by (10). Using the law of total probability, we rewrite (11) as
\begin{equation}\label{equa12}
\begin{split}
P_{{\textrm{out}}}  = & \Pr ( {\left. {{\cal{D}} = \emptyset, H_0 } \right|\hat H = H_0 } ) \\
&+ \Pr ( {\left. {{\cal{D}} = \emptyset, H_1 } \right|\hat H = H_0 } )\\
&+ \sum\limits_{n = 1}^{2^N  - 1} {\Pr ( { {C_d  < R,{\cal{D}} = {\cal{D}}_n, H_0 } |\hat H = H_0 } )}\\
&+ \sum\limits_{n = 1}^{2^N  - 1} {\Pr ( { {C_d  < R,{\cal{D}} = {\cal{D}}_n, H_1 } |\hat H = H_0 } )},
\end{split}
\end{equation}
which can be further expressed as
\begin{equation}\label{equa13}
\begin{split}
P_{{\textrm{out}}} = & \Pr ( {\left. {{\cal{D}} = \emptyset } \right|H_0, \hat H = H_0 } )\Pr (H_0 | \hat H = H_0) \\
&+ \Pr ( {\left. {{\cal{D}} = \emptyset} \right|H_1, \hat H = H_0 } )\Pr (H_1 | \hat H = H_0)\\
&+ \sum\limits_{n = 1}^{2^N  - 1} {\Pr ( { {C_d  < R,{\cal{D}} = {\cal{D}}_n } | H_0, \hat H = H_0 } )}\\
&\quad\quad\quad\times \Pr (H_0 | \hat H = H_0)\\
&+ \sum\limits_{n = 1}^{2^N  - 1} {\Pr ( { {C_d  < R,{\cal{D}} = {\cal{D}}_n } | H_1, \hat H = H_0 } )}\\
&\quad\quad\quad\times \Pr (H_1 | \hat H = H_0),
\end{split}
\end{equation}
where $\Pr ( {H_0 | {\hat H  = H_0 } } )$ and $\Pr ( {H_1 | {\hat H  = H_0 } } )$ can be computed by using Bayes' theorem as
\begin{equation}\label{equa14}
\begin{split}
 \Pr ( {H_0 | {\hat H  = H_0 } } ) = & \frac{{\Pr ( {\hat H  = H_0 | {H_0 } } )\Pr ( {H_0 } )}}{{\sum\limits_{i \in \{0,1\}} {\Pr ( {\hat H  = H_0 | {H_i } } )\Pr ( {H_i } )} }} \\
  = &\frac{{P_0 P_d}}{{P_0 P_d + (1 - P_0 )P_f}} \buildrel \Delta   \over = \pi _0,
 \end{split}
\end{equation}
and
\begin{equation}\label{equa15}
\Pr ( {H_1 | {\hat H  = H_0 } } ) = \frac{{(1 - P_0 )P_f }}{{P_0 P_d  + (1 - P_0 )P_f}}\buildrel \Delta \over = \pi _1,
\end{equation}
where $P_0  = \Pr ( {H_0 } )$ is the probability that the licensed spectrum becomes unoccupied by PT, while ${{{P}}_{{d}}} = \Pr ({\hat H} = {H_0}|{H_0})$ and ${{{P}}_{{f}}} = \Pr ({\hat H} = {H_0}|{H_1})$ are the probability of detection and false alarm of the presence of a spectrum hole, respectively. For notational convenience, we denote $\pi _0  = \Pr ( {H_0 | {\hat H  = H_0 } } )$, $\pi _1  = \Pr ( {H_1 | {\hat H  = H_0 } } )$ and $\Delta  = \frac{{2^{2R}  - 1}}{{\gamma _s }}$. Thus, using (2)-(5), (10), (14) and (15), we can rewrite (13) as
\begin{equation}\label{equa16}
\begin{split}
 P_{{\textrm{out}}}  &= \pi _0 \prod\limits_{i = 1}^N {\Pr ( {|h_{si} |^2  < \Delta } )}  \\
&+ \pi _1 \prod\limits_{i = 1}^N {\Pr ( {|h_{si} |^2  < \Delta |h_{pi} |^2 \gamma _p  + \Delta } )}  \\
&+ \pi _0 \sum\limits_{n = 1}^{2^N  - 1} {\prod\limits_{i \in {\cal{D}}_n } {\Pr ( {|h_{si} |^2  > \Delta } )}}\\
&\quad\quad\quad\times \prod\limits_{j \in \bar {\cal{D}}_n } {\Pr ( {|h_{sj} |^2  < \Delta } )} \\
&\quad\quad\quad\times \Pr ( {\sum\limits_{i \in {\cal{D}}_n } {|h_{id} |^2 }  < \Delta } )  \\
&+ \pi _1 \sum\limits_{n = 1}^{2^N  - 1} {\prod\limits_{i \in {\cal{D}}_n } {\Pr ( {|h_{si} |^2  > \Delta |h_{pi} |^2 \gamma _p  + \Delta } )}} \\
&\quad\quad\quad\times \prod\limits_{j \in \bar {\cal{D}}_n } {\Pr ( {|h_{sj} |^2  < \Delta |h_{pj} |^2 \gamma _p  + \Delta } )}   \\
&\quad\quad\quad \times \Pr ( {\sum\limits_{i \in {\cal{D}}_n } {|h_{id} |^2 }  < \gamma _p \Delta |h_{pd} |^2  + \Delta } ) \\
\end{split}
\end{equation}
Since random variables $|{h_{si}}{|^2}$ and $|{h_{pi}}{|^2}$ are independently and exponentially distributed with respective means of $\sigma _{si}^{{\rm{  }}2}$ and $\sigma _{pi}^{{\rm{  }}2}$, we readily obtain
\begin{equation}\label{equa17}
\Pr (|{h_{si}}{|^2} < \Delta ) = 1 - \exp ( - \frac{\Delta }{{\sigma _{si}^2}}),
\end{equation}
and
\begin{equation}\label{equa18}
\Pr (|{h_{si}}{|^2} - |{h_{pi}}{|^2} {\gamma _p}\Delta < \Delta ) = 1 - \frac{{\sigma _{si}^2}}{{\sigma _{pi}^2{\gamma _p}\Delta  + \sigma _{si}^2}}\exp ( - \frac{\Delta }{{\sigma _{si}^2}}).
\end{equation}
It is challenging to obtain general closed-form expressions of $\Pr (\sum\limits_{i \in {\cal{D}}_n } {|h_{id} |^2 }  < \Delta )$ and $\Pr ( {\sum\limits_{i \in {\cal{D}}_n } {|h_{id} |^2 }  < \gamma _p \Delta |h_{pd} |^2  + \Delta } )$. For simplicity, we here assume that the fading coefficients of all relay-SD channels $|h_{id}|^2$ are independent identically distributed (i.i.d.) random variables with the same average channel gain denoted by $\sigma^2_{d}=E(|h_{id}|^2)$. Hence, we can obtain
\begin{equation}\label{equa19}
\Pr (\sum\limits_{i \in {\cal{D}}_n } {|h_{id} |^2 }  < \Delta ) = \Gamma ( {\frac{{\Delta}}{{ \sigma _d^2 }},|{\cal{D}}_n |} ),
\end{equation}
and
\begin{equation}\label{equa20}
\begin{split}
& \Pr (\sum\limits_{i \in {\cal D}_n } {|h_{id} |^2 }  < \gamma _p \Delta |h_{pd} |^2  + \Delta ) =\Gamma (\frac{{\Delta }}{{\sigma _d^2 }},|{\cal D}_n |)\\
&\quad \quad + \frac{{[1 - \Gamma (\Delta \sigma _d^{ - 2}  + \sigma _{pd}^{ - 2} \gamma _p^{ - 1} ,|{\cal{D}}_n |)]}}{{(1 + \sigma _d^2 \sigma _{pd}^{ - 2} \gamma _p^{ - 1} \Delta ^{ - 1} )^{|{\cal{D}}_n |} }}e^{1/(\sigma _{pd}^2 \gamma _p )}, \\
 \end{split}
\end{equation}
where $\Gamma (x,k) = \int_0^x {\frac{{t^{k - 1} }}{{\Gamma (k)}}e^{ - t} dt}$ is known as the incomplete Gamma function. Substituting (17)-(20) into (16) readily yields a closed-form expression of the outage probability for the proposed multi-relay selection scheme.

\section{Numerical Results and Discussions}
This section presents the numerical performance evaluation of proposed multi-relay selection scheme in terms of the outage probability. Also, the outage performance of the conventional direct transmission and best-relay selection is evaluated for comparison purposes. In Fig. 2, we depict the outage probability versus the secondary transmit power $\gamma_s$ of the conventional direct transmission and best-relay selection as well as the proposed multi-relay selection schemes for different $(P_d,P_f)$. It can be observed from Fig. 2 that as the spectrum sensing requirement is improved from $(P_d,P_f)=(0.65,0.35)$ to $(P_d,P_f)=(0.95,0.05)$, the outage probabilities of the three schemes decrease accordingly. This is because that with an improved spectrum sensing requirement, a spectrum hole would be detected more accurately and thus less interference occurs between the primary and secondary networks, resulting in a decreased outage probability for the secondary transmissions. One can also see from Fig. 3 that for both $(P_d,P_f)=(0.95,0.05)$ and $(P_d,P_f)=(0.65,0.35)$, the outage performance of the proposed multi-relay selection is better than that of the conventional direct transmission and best-relay selection schemes. Additionally, the theoretical and simulation results match well, confirming the correctness of the closed-form outage analysis.
\begin{figure}
  \centering
  {\includegraphics[scale=0.55]{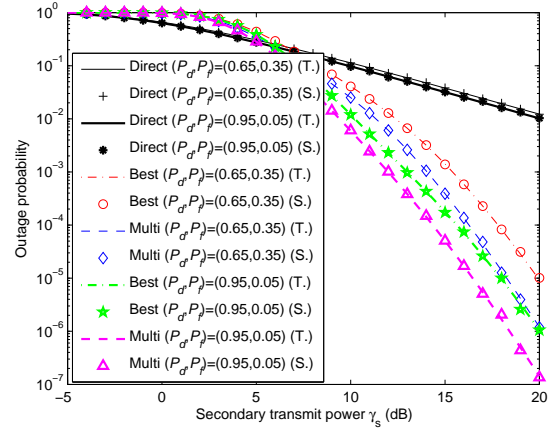}\\
  \caption{Outage probability versus the secondary transmit power $\gamma_s$ of the conventional direct transmission and best-relay selection as well as the proposed multi-relay selection schemes for different $(P_d,P_f)$ with $P_0=0.8$, $\gamma_p=10{\textrm{dB}}$, $R=1{\textrm{bit/s/Hz}}$, $N=6$, $\sigma^2_{sd}=\sigma^2_{si}=\sigma^2_{id}=1$ and $\sigma^2_{pd}=\sigma^2_{pi}=0.2$, where the shorthand `T.' and `S.' stand for the theoretical and simulation results, respectively.}\label{Fig2}}
\end{figure}

Fig. 3 shows the outage probability versus the secondary transmit power $\gamma_s$ of the conventional direct transmission and best-relay selection as well as the proposed multi-relay selection schemes for different $N$. It is shown in Fig. 3 that for both $N=4$ and $N=6$, the best-relay selection and the multi-relay selection schemes both perform better than the direct transmission in terms of the outage probability. Moreover, as the number of relays increases from $N=4$ to $6$, the outage performance of the best-relay selection and multi-relay selection both improves significantly, demonstrating the outage benefits of exploiting relay selection for assisting the secondary transmissions. Additionally, Fig. 3 also shows the outage advantage of proposed multi-relay selection over conventional direct transmission and best-relay selection schemes.
\begin{figure}
  \centering
  {\includegraphics[scale=0.55]{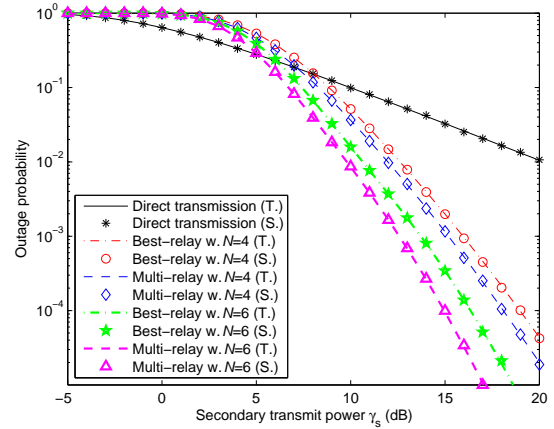}\\
  \caption{Outage probability versus the secondary transmit power $\gamma_s$ of the conventional direct transmission and best-relay selection as well as the proposed multi-relay selection schemes for different $N$ with $P_0=0.8$, $\gamma_p=10{\textrm{dB}}$, $R=1{\textrm{bit/s/Hz}}$, $(P_d,P_f)=(0.9,0.1)$, $\sigma^2_{sd}=\sigma^2_{si}=\sigma^2_{id}=1$ and $\sigma^2_{pd}=\sigma^2_{pi}=0.2$.}\label{Fig3}}
\end{figure}

\section{Conclusions}
In this paper, we proposed a multi-relay selection scheme for a cognitive radio network comprised of a ST, a SD and multiple DF relays, where multiple relays are selected to forward the ST's transmissions to SD. We derived a closed-form expression of the outage probability for the proposed multi-relay selection assisted secondary transmissions. The numerical outage results of the proposed multi-relay selection were provided and the conventional direct transmission as well as the best-relay selection were also considered for comparison purposes. It was shown that the proposed multi-relay selection outperforms the conventional direct transmission and best-relay selection in terms of the outage probability.

\appendices

\end{document}